\documentclass{mem}
\usepackage{natbib}\usepackage{txfonts}\usepackage{balance}
\usepackage{graphicx}
\usepackage{epstopdf}
\usepackage[a4paper]{hyperref}
\idline{75}{282}
\begin{document}
\title{The solar continuum intensity distribution: Settling the conflict between observations and simulations}

   \subtitle{}

\author{S. \, Wedemeyer-B\"ohm\inst{1,2} 
\and 
L. \, Rouppe van der Voort\inst{1}}

\offprints{S. Wedemeyer-B\"ohm}

\institute{
Institute of Theoretical Astrophysics, University of Oslo,
  P.O. Box 1029 Blindern, N-0315 Oslo, Norway
  \and
  Center of Mathematics for Applications (CMA), University of Oslo,
  Box 1053 Blindern, NÐ0316 Oslo, Norway
\email{sven.wedemeyer-bohm@astro.uio.no}
}

\authorrunning{Wedemeyer-B\"ohm  \& Rouppe van der Voort}

\titlerunning{The solar continuum intensity distribution}

\abstract{
For many years, there seemed to be significant differences between the 
continuum intensity distributions derived from observations and simulations 
of the solar photosphere. In order to settle the discussion on these apparent 
discrepancies, we present a detailed comparison between simulations and 
seeing-free observations that takes into account the crucial influence of 
instrumental image degradation. 
We use a set of images of quiet Sun granulation taken in the blue, green and red 
continuum bands of the Broadband Filter Imager of the Solar Optical Telescope (SOT) 
onboard Hinode. 
The images are deconvolved with Point Spread Functions (PSF) that account for non-ideal
contributions due to instrumental stray-light and imperfections.  
In addition, synthetic intensity images are degraded with the corresponding PSFs. 
The results are compared with respect to spatial power spectra, intensity histograms, 
and the centre-to-limb variation of the intensity contrast.
The observational findings are well matched with corresponding synthetic observables 
from three-dimensional radiation (magneto-)hydrodynamic simulations. 
We conclude that the intensity contrast of the solar continuum intensity is higher 
than usually derived from ground-based observations and is well reproduced by modern 
numerical simulations. 
Properly accounting for image degradation effects is of crucial importance for 
comparisons between observations and numerical models. It finally settles the 
traditionally perceived conflict between observations 
and simulations. 
}
\maketitle%{}

%-----------------------------------------------------------------------
%-----------------------------------------------------------------------
%-----------------------------------------------------------------------
\section{Introduction}

%-----------------------------------------------------------------------
\begin{figure*}[]
\centering
\resizebox{12cm}{!}{\includegraphics[clip=true]{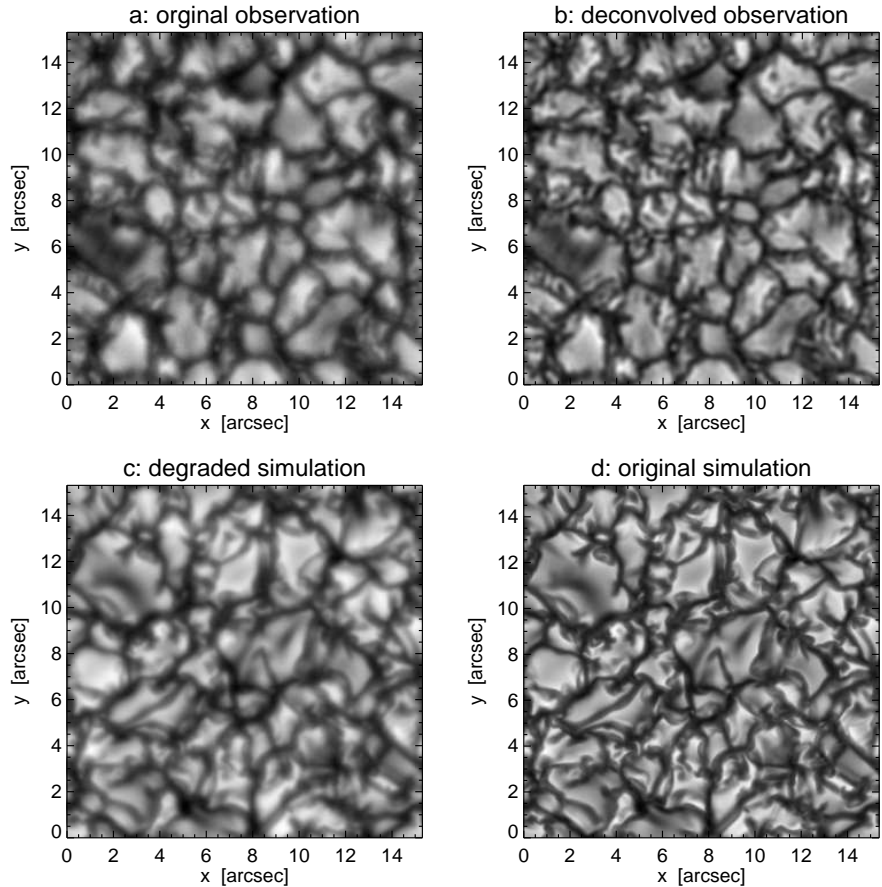}}
\caption{\footnotesize
Comparison of exemplary intensity maps:  
\textbf{a)}~original observation (close-up), 
\textbf{b)}~deconvolved observation, 
\textbf{c)}~degraded simulation, 
\textbf{d)}~original simulation. }
\label{fig:wbvdv_maps}
\end{figure*}
%-----------------------------------------------------------------------
%-----------------------------------------------------------------------
\begin{figure}[t]
\resizebox{\hsize}{!}{\includegraphics[clip=true]{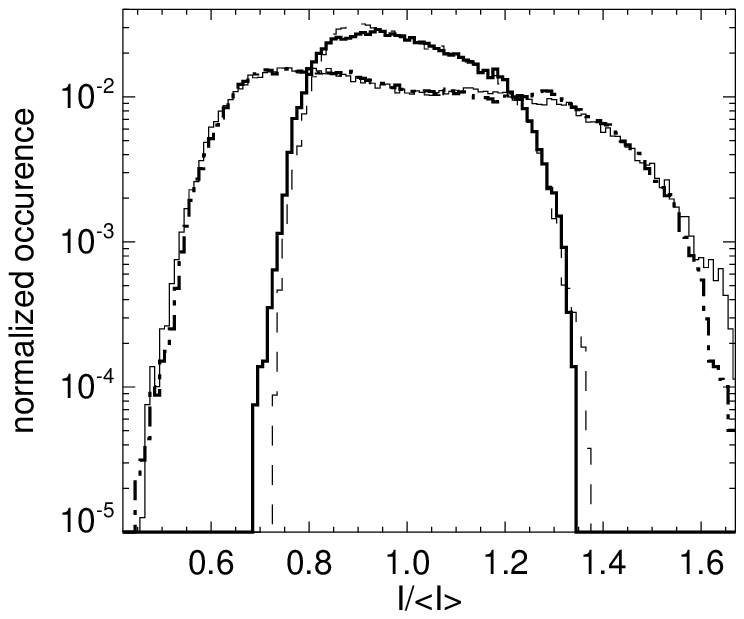}}
\caption{
\footnotesize
Intensity distribution of the exemplary intensity images: 
original observation (thin dashed), 
deconvolved observation (thin solid), 
degraded simulation (thick solid), 
original simulation (thick dot-dashed).}
\label{fig:wbvdv_hist}
\end{figure}
%-----------------------------------------------------------------------

The contrast of the continuum intensity originating from the low solar 
photosphere represents one of many  diagnostic tests that are needed to 
check the realism of numerical models of the solar atmosphere. 
\citep[see, e.g.,][]{1975A&A....45..167D}. 
In the past, there seemed to be a disturbing discrepancy in granulation 
contrast measurements derived from observations of the quiet Sun and 
from three-dimensional numerical models \citep[see, e.g.,][]{1984ssdp.conf..174N}. 
The historic evolution of the contrast measurements
is illustrated in Fig.~1 by \citet{2008PhST..133a4016K} 
for a continuum wavelength of 500\,nm.  
Simulations produce values between 18\,\% and 27\,\%, whereas 
uncorrected observations have contrasts of only a few percent points. 
See also, e.g., \citet{1990ARA&A..28..263S} and \citet{2000ApJ...538..940S}. 

This apparent discrepancy is now proven to be caused by the problematic 
correction of the observed intensity images. 
Observations have to be corrected for degradation due to instrumental effects 
and -- in case of ground-based observations -- the terrestrial atmosphere
(``seeing'').  
Unfortunately, the properties of the degradation are difficult to be 
determined and therefore often only poorly known. 
The result is an often incomplete correction and a correspondingly too 
low empirical granulation contrast. 

This short article summarises some of the main points from a recent 
more extensive study that we published in \citet{2009arXiv0905.0705W}. 
By using space-borne observations with the Solar Optical Telescope    
\citep[SOT,][]{2008SoPh..249..167T,
2008ASPC..397....5I,
2008SoPh..249..197S,
2008SoPh..249..221S}
onboard the Hinode spacecraft \citep{2007SoPh..243....3K}, 
only the instrumental degradation had to be corrected for, which enabled 
a meaningful comparison of observations and state-of-the-art numerical 
simulations of the solar photosphere. 

% ================================================================================
\section{Observations}

We selected 584~images taken with the Broadband Filter Imager (BFI) of the 
Solar Optical Telescope (SOT). 
The blue (450\,nm), green (555\,nm), and red (668\,nm) wavelength bands 
have been considered.
Only quiet Sun regions at positions from disc-centre to the limb were chosen. 

SOT has a spatial resolution of $\sim 0\,\farcs2 - 0\,\farcs3$, which is clearly 
sufficient to resolve the solar granulation pattern.  
In a precursory study, statistically representative point spread functions (PSFs) 
were constructed for each of the channels considered here
\citep{2008A&A...487..399W}.  
For those, a total of ~70 images of eclipse and Mercury transit observations 
were employed. 
The PSFs included the effect of the SOT aperture and an approximation for non-ideal 
stray-light contributions. 

% ================================================================================
\section{Synthetic intensity}

Most of our analysis was based on three snapshots taken from a recent 3D radiation 
hydrodynamic simulation, which was carried out  
by \citet{2007IAUS..239...36S} with \mbox{CO$^5$BOLD} \citep{2008asd..soft...36F}.  
For comparison, additional simulations by \citet{2006ASPC..354..345S}, 
\citet{2004A&A...414.1121W},    
and  \citet{1998ApJ...499..914S} were used. 
The individual models differ in horizontal grid spacing and extent, the 
frequency-dependence of radiative transfer, and the inclusion of magnetic 
fields. 

The intensity synthesis code 
Linfor3D (see \mbox{http://www.aip.de/$\sim$mst/linfor3D$\_$main.html)}
was used to calculate synthetic continuum intensity maps 
for each wavelength band and all simulation snapshots for 
heliocentric positions from disc-centre to limb with an increment 
of $\Delta\mu =0.05$. 

% ================================================================================
\section{Comparison of observed and synthetic images}
\label{sec:compwsim}

We now compare exemplary intensity maps for the blue channel. 
Next a close-up from the original filtergram, we show the corrected intensity 
image in Fig.~\ref{fig:wbvdv_maps}. 
The image degradation due to the instrument was corrected for by deconvolution 
with the detailed PSF.  
In this example, the granulation contrast increases from 12.4\,\% to 
25.4\,\%. 
This compares to 25.0\,\% for the original simulation snapshot. 
Degrading the latter by applying the PSF decreases the synthetic contrast to 
12.9\,\%. 
A more detailed mean of comparison are the intensity histograms displayed in 
Fig.~\ref{fig:wbvdv_hist}. 
The original simulation has a much broader distribution than the 
original observation and exhibits two distinct peaks at darker and brighter 
than average intensity values, which is not discernible from the narrow 
observed distribution. 
Correcting for the influence of instrumental image degradation, however, produces 
a close match of observed and synthetic intensity distribution. 

% ================================================================================
\section{Conclusions}

We conclude that the traditionally perceived conflict between observations 
and simulations in terms of granulation contrast can be dismissed. 
Modern simulations are obviously sufficiently realistic to reproduce the main 
characteristics of the (lower) solar photosphere in quiet Sun regions. 
It is however important to note that both sides have small but noticeable 
intrinsic variations, e.g., caused by selection of the field of view. 
One can therefore not expect an exact match of the contrast numbers. 

The granulation contrast as a single number is a poor test of the 
realism of simulations. 
Evaluating a complex phenomenon like the solar surface convection 
requires more substantial tests. In this respect it is essential to note 
that also the centre-to-limb variation of the contrast and 
the power spectral density of the continuum intensity from observations 
and simulations are in good agreement. 

Detailed PSFs with reliable estimates of the stray-light contributions are 
crucial for quantitative comparisons as shown here. 
Ideally one would measure the PSF exactly simultaneous to recording an 
intensity image. 
As this is essentially not possible, a robust correction for the influence of 
image degradation requires significant statistics employing a large 
number of intensity images as done by \citet{2008A&A...487..399W} and 
\citet{2009arXiv0905.0705W}. 
Studies using PSFs and intensity maps based on a single or a few images only 
are potentially misleading. 

The next step towards realistic models concerns the refinement of the 
thermal structure and velocity field above the lower photosphere. 
An obvious way to test it are detailed comparisons of spectral lines 
as they are currently employed in the context of abundance determinations 
\citep[see, e.g.,][]{2000A&A...359..729A,2008A&A...488.1031C}. 

%-----------------------------------------------------------------------
\begin{acknowledgements}
SWB thanks the organisers of Joint Discussion 10 held at IAU General 
Assembly in Rio de Janeiro, Brazil, in 2009. 
This work was supported by the Research Council of Norway, grant 
170935/V30, and a Marie Curie Intra-European Fellowship of the European Commission 
(6th Framework Programme). 
\end{acknowledgements}

%-----------------------------------------------------------------------
\bibliographystyle{aa}

\begin{thebibliography}{}

\bibitem[{{Asplund} {et~al.}(2000){Asplund}, {Nordlund}, {Trampedach}, {Allende
  Prieto}, \& {Stein}}]{2000A&A...359..729A}
{Asplund}, M., {Nordlund}, {\AA}., {Trampedach}, R., {Allende Prieto}, C., \&
  {Stein}, R.~F. 2000, \aap, 359, 729

\bibitem[{{Caffau} {et~al.}(2008){Caffau}, {Ludwig}, {Steffen}, {Ayres},
  {Bonifacio}, {Cayrel}, {Freytag}, \& {Plez}}]{2008A&A...488.1031C}
{Caffau}, E., {Ludwig}, H.-G., {Steffen}, M., {et~al.} 2008, \aap, 488, 1031

\bibitem[{{Deubner} \& {Mattig}(1975)}]{1975A&A....45..167D}
{Deubner}, F.~L. \& {Mattig}, W. 1975, \aap, 45, 167

\bibitem[{{Freytag} {et~al.}(2008){Freytag}, {Steffen}, {Ludwig}, \&
  {Wedemeyer-Boehm}}]{2008asd..soft...36F}
{Freytag}, B., {Steffen}, M., {Ludwig}, H.-G., \& {Wedemeyer-Boehm}, S. 2008,
  Astrophysics Software Database, 36

\bibitem[{{Ichimoto} {et~al.}(2008){Ichimoto}, {Katsukawa}, {Tarbell}, {Shine},
  {Hoffmann}, {Berger}, {Cruz}, {Suematsu}, {Tsuneta}, {Shimizu}, \&
  {Lites}}]{2008ASPC..397....5I}
{Ichimoto}, K., {Katsukawa}, Y., {Tarbell}, T., {et~al.} 2008, in Astronomical
  Society of the Pacific Conference Series, Vol. 397, First Results From
  Hinode, ed. S.~A. {Matthews}, J.~M. {Davis}, \& L.~K. {Harra}, 5

\bibitem[{{Kiselman}(2008)}]{2008PhST..133a4016K}
{Kiselman}, D. 2008, Physica Scripta Volume T, 133, 014016

\bibitem[{{Kosugi} {et~al.}(2007){Kosugi}, {Matsuzaki}, {Sakao}, {Shimizu},
  {Sone}, {Tachikawa}, {Hashimoto}, {Minesugi}, {Ohnishi}, {Yamada}, {Tsuneta},
  {Hara}, {Ichimoto}, {Suematsu}, {Shimojo}, {Watanabe}, {Shimada}, {Davis},
  {Hill}, {Owens}, {Title}, {Culhane}, {Harra}, {Doschek}, \&
  {Golub}}]{2007SoPh..243....3K}
{Kosugi}, T., {Matsuzaki}, K., {Sakao}, T., {et~al.} 2007, \solphys, 243, 3

\bibitem[{{Nordlund}(1984)}]{1984ssdp.conf..174N}
{Nordlund}, A. 1984, in Small-Scale Dynamical Processes in Quiet Stellar
  Atmospheres, ed. S.~L. {Keil}, 174

\bibitem[{{S{\'a}nchez Cuberes} {et~al.}(2000){S{\'a}nchez Cuberes}, {Bonet},
  {V{\'a}zquez}, \& {Wittmann}}]{2000ApJ...538..940S}
{S{\'a}nchez Cuberes}, M., {Bonet}, J.~A., {V{\'a}zquez}, M., \& {Wittmann},
  A.~D. 2000, \apj, 538, 940

\bibitem[{{Schaffenberger} {et~al.}(2006){Schaffenberger},
  {Wedemeyer-B{\"o}hm}, {Steiner}, \& {Freytag}}]{2006ASPC..354..345S}
{Schaffenberger}, W., {Wedemeyer-B{\"o}hm}, S., {Steiner}, O., \& {Freytag}, B.
  2006, in ASP Conf. Series, Vol. 354,
  Solar MHD Theory and Observations: A High Spatial Resolution Perspective, ed.
  J.~{Leibacher}, R.~F. {Stein}, \& H.~{Uitenbroek}, 345

\bibitem[{{Shimizu} {et~al.}(2008){Shimizu}, {Nagata}, {Tsuneta}, {Tarbell},
  {Edwards}, {Shine}, {Hoffmann}, {Thomas}, {Sour}, {Rehse}, {Ito},
  {Kashiwagi}, {Tabata}, {Kodeki}, {Nagase}, {Matsuzaki}, {Kobayashi},
  {Ichimoto}, \& {Suematsu}}]{2008SoPh..249..221S}
{Shimizu}, T., {Nagata}, S., {Tsuneta}, S., {et~al.} 2008, \solphys, 249, 221

\bibitem[{{Spruit} {et~al.}(1990){Spruit}, {Nordlund}, \&
  {Title}}]{1990ARA&A..28..263S}
{Spruit}, H.~C., {Nordlund}, A., \& {Title}, A.~M. 1990, \araa, 28, 263

\bibitem[{{Steffen}(2007)}]{2007IAUS..239...36S}
{Steffen}, M. 2007, in IAU Symposium, Vol. 239, IAU Symposium, ed. F.~{Kupka},
  I.~{Roxburgh}, \& K.~{Chan}, 36--43

\bibitem[{{Stein} \& {Nordlund}(1998)}]{1998ApJ...499..914S}
{Stein}, R.~F. \& {Nordlund}, A. 1998, \apj, 499, 914

\bibitem[{{Suematsu} {et~al.}(2008){Suematsu}, {Tsuneta}, {Ichimoto},
  {Shimizu}, {Otsubo}, {Katsukawa}, {Nakagiri}, {Noguchi}, {Tamura}, {Kato},
  {Hara}, {Kubo}, {Mikami}, {Saito}, {Matsushita}, {Kawaguchi}, {Nakaoji},
  {Nagae}, {Shimada}, {Takeyama}, \& {Yamamuro}}]{2008SoPh..249..197S}
{Suematsu}, Y., {Tsuneta}, S., {Ichimoto}, K., {et~al.} 2008, \solphys, 249,
  197

\bibitem[{{Tsuneta} {et~al.}(2008){Tsuneta}, {Ichimoto}, {Katsukawa}, {Nagata},
  {Otsubo}, {Shimizu}, {Suematsu}, {Nakagiri}, {Noguchi}, {Tarbell}, {Title},
  {Shine}, {Rosenberg}, {Hoffmann}, {Jurcevich}, {Kushner}, {Levay}, {Lites},
  {Elmore}, {Matsushita}, {Kawaguchi}, {Saito}, {Mikami}, {Hill}, \&
  {Owens}}]{2008SoPh..249..167T}
{Tsuneta}, S., {Ichimoto}, K., {Katsukawa}, Y., {et~al.} 2008, \solphys, 249,
  167

\bibitem[{{Wedemeyer} {et~al.}(2004){Wedemeyer}, {Freytag}, {Steffen},
  {Ludwig}, \& {Holweger}}]{2004A&A...414.1121W}
{Wedemeyer}, S., {Freytag}, B., {Steffen}, M., {Ludwig}, H.-G., \& {Holweger},
  H. 2004, \aap, 414, 1121

\bibitem[{{Wedemeyer-B{\"o}hm}(2008)}]{2008A&A...487..399W}
{Wedemeyer-B{\"o}hm}, S. 2008, \aap, 487, 399

\bibitem[{{Wedemeyer-B{\"o}hm} \& {Rouppe van der
  Voort}(2009)}]{2009arXiv0905.0705W}
{Wedemeyer-B{\"o}hm}, S. \& {Rouppe van der Voort}, L. 2009, \aap, 503, 225

\end{thebibliography}

\end{document}